\documentclass[preprint2]{aastex}
\def\stacksymbols #1#2#3#4{\def\theguybelow{#2}
	\def\verticalposition{\lower#3pt}
	\def\spacingwithinsymbol{\baselineskip0pt\lineskip#4pt}
	\mathrel{\mathpalette\intermediary#1}}
\def\intermediary #1#2{\verticalposition\vbox{\spacingwithinsymbol
	\everycr={}\tabskip0pt
	\halign{$\mathsurround0pt#1\hfil##\hfil$\crcr#2\crcr
		\theguybelow\crcr}}}
\def\lta{\stacksymbols{<}{\sim}{2.5}{.2}}
\def\gta{\stacksymbols{>}{\sim}{3}{.5}}

\begin{document}

\title{WHY ARE ROTATING 
ELLIPTICAL GALAXIES LESS ELLIPTICAL AT X-RAY FREQUENCIES?}

\author{Fabrizio Brighenti$^{1,2}$ and William G. Mathews$^1$}

\affil{$^1$University of California Observatories/Lick Observatory,
Board of Studies in Astronomy and Astrophysics,
University of California, Santa Cruz, CA 95064\\
mathews@lick.ucsc.edu}

\affil{$^2$Dipartimento di Astronomia,
Universit\`a di Bologna,
via Ranzani 1,
Bologna 40127, Italy\\
brighenti@bo.astro.it}






\vskip .2in

\begin{abstract}
If mass and angular momentum are conserved in the cooling flows 
associated with luminous, slowly rotating 
elliptical galaxies, the flow cools onto 
extended, massive disks of rotationally supported cold gas 
in the equatorial plane. 
As the hot interstellar gas  
approaches the disks, its density and 
thermal X-ray emission increase, 
resulting in X-ray images that are 
considerably flattened toward the equatorial plane out to an 
optical effective radius or beyond.
Remarkably, the flattening of X-ray images due to rotation is 
very small or absent 
at the spatial resolution currently available to 
X-ray observations.
This is strong evidence that mass and angular momentum
are not in fact conserved.
In particular, if cooling flows are  
depleted by localized radiative cooling at numerous 
sites distributed throughout the 
flows, disks of cooled gas do not form and the X-ray images  
appear nearly circular.
Even in this case, however, 
the spatial distribution of the cooled gas, 
and any young stars that may have formed from this gas, 
would be decidedly flatter than 
the old stellar population; if the young stars are 
optically luminous, the Balmer lines they contribute 
to the stellar spectra 
should be more elliptical than the total stellar light.
In principle, X-ray images of galactic cooling 
flows can also be circularized 
by the turbulent diffusion of angular momentum 
away from the axis of rotation.
But the effective viscosity of known processes 
-- stellar mass loss, supernovae, cooling site evolution, etc. -- 
is insufficient to circularize 
the X-ray images appreciably. 
Radial gradients in the interstellar iron abundance in elliptical
galaxies similar to those observed 
are unaffected by the expected level of interstellar turbulence
since these gradients are quickly 
re-established by Type Ia supernovae.

\end{abstract}


\clearpage

\section{INTRODUCTION}

Cooling flows in massive elliptical galaxies 
are expected to rotate since most of the interstellar
gas within an optical 
effective radius $R_e$ is produced by mass loss 
from the slowly rotating stellar systems.
If the cooling flow proceeds inward 
conserving angular momentum, the interstellar gas 
ultimately flows toward a large disk ($R_{disk} \gta R_e$) 
and spins up to the local equatorial circular velocity,
$\sim 400$ km s$^{-1}$
(Kley \& Mathews 1995; Brighenti \& Mathews 1996, 
subsequently referred to as Paper I).
As interstellar gas approaches the disk,
the X-ray images are significantly flattened 
toward the equatorial plane on scales $\sim R_e - 3R_e$
when viewed perpendicular to the axis of rotation. 
For bright elliptical galaxies at distance 17 Mpc in the Virgo
cluster -- NGC 4472, NGC 4636, and NGC 4649 -- $R_e \sim 7~{\rm
kpc}~\sim 1.4$'.  However, {\it Einstein} X-ray images show no
evidence of rotational flattening either at the resolution of the IPC
($\sim 55'' \sim 4.5$ kpc; Trinchieri et al. 1986) or at the much
higher resolution or HRI ($\sim 6'' \sim 0.49$ kpc; Fabbiano, Kim \&
Trinchieri 1992).
In a series of papers Buote \& Canizares have detected 
large scale X-ray 
flattening in several massive elliptical galaxies which they
attribute to a flattened dark matter potential, but 
the highly flattened inner isophotes 
anticipated from momentum-conserving cooling flows 
were not evident (see Buote \& Canizares 1998 for a 
review). 
More recently, Hanlan \& Bregman (1999) have found little 
or no rotationally enhanced 
X-ray ellipticity in ROSAT HRI and PSPC images of 
several bright elliptical galaxies.
The implications 
of these observations are not 
widely appreciated among the community of astronomers 
interested in early type galaxies.

Massive, X-ray luminous elliptical galaxies have two 
principal sources of hot 
interstellar gas: (1) ejection of stellar envelopes from 
an evolving population of old stars and (2) inflow of 
distant circumgalactic gas which 
accumulates in the galactic halo either by tidal exchange 
or secondary infall (Brighenti \& Mathews 1999a). 
If the circumgalactic gas has a lower specific angular momentum 
than the stars, 
the X-ray flattening would be lessened 
as inflowing circumgalactic gas mixes with 
gas lost from the rotating stellar system within 
several $R_e$.
But it is unlikely that X-ray disks 
can be eliminated by this means 
since most of the gas in the inner galaxy is produced by 
stellar mass loss and has not flowed in from the 
galactic halo region.
Our recent models for the large elliptical galaxy NGC 4472
require that at least 60 - 70 percent 
of the hot gas in $r \lta R_e$ 
originates from stellar mass loss. 
This rate of stellar mass ejection is expected from 
normal stellar evolution and is required 
to explain the radial variation of interstellar
temperature and metallicity typically observed 
in massive elliptical galaxies (Brighenti \& Mathews 1999a).
In addition,
it is unlikely that the absence of observed X-ray flattening 
can be understood simply by the inflow of halo gas 
into the optical centers of elliptical galaxies since 
it is plausible that the extended halo gas has an even 
{\it larger} specific angular momentum than the stars. 
N-body simulations of galaxy formation typically 
produce elliptical galaxies and galaxy groups
with increasing specific angular momentum at large radii
(e.g. Barnes \& Efstathiou 1987; Quinn \& Zureck 1988).
But the intrinsic angular momentum 
in distant halo gas is uncertain; 
gas acquired by tidal disruption is expected 
to have significant rotation, but 
secondary infalling gas that arrives after most
mergers have occurred may have less net rotation than the 
stellar system.

However, there is no doubt that the stellar systems in 
most large elliptical galaxies 
and the gas that they expel are rotating.
Although the stellar systems in
luminous elliptical galaxies (with $L_B \gta 3 \times 10^{10}$ 
$L_{B,\odot}$) are not flattened
by rotation, their rotation about the minor axis is not
small, typically 50 - 100 km s$^{-1}$
(e.g. Binney, Davies, \& Illingworth 1990), and 
the gas they expel must spin up further as it moves inward.

These general considerations indicate that the net angular 
momentum of gas ejected from stars must be removed 
to circularize the X-ray images. 
This can be accomplished either 
by removing gas from the flow 
or by transporting angular momentum 
away from the axis of rotation by viscous interactions. 
In this paper we explore the importance of these two 
processes and determine 
their relative influence on the X-ray images.
The first process, localized dropout cooling of 
interstellar gas in regions of low specific entropy, 
is also required to limit the masses of central black holes 
(or dark stellar nuclei) in elliptical galaxies 
to their observed values. 
Central black holes would be about ten times more massive 
if the cooling flow proceeded all the way to the galactic 
center with constant mass flux 
(Brighenti \& Mathews 1999b).
The presence of distributed optical emission lines from 
cooling gas also argues for cooling dropout 
(Mathews \& Brighenti 1999a). 
However, as we show below, the strongest argument for the 
presence of localized mass dropout in the 
interstellar gas may be the circularization of X-ray images 
that results.

We also explore the possibility that 
interstellar viscosity transports angular momentum outward, 
reducing the rotation and X-ray flattening in the inner galaxy. 
An effective viscosity can arise from 
several natural sources: 
plasma viscosity,  
interstellar ``turbulent'' viscosity 
driven by stellar mass loss or Type Ia supernovae, 
the viscosity implicit in regions of localized interstellar cooling 
as they sink in the galactic potential, 
and turbulent viscosity 
that may develop from shear instabilities in the flow.
On several occasions Nulsen and Fabian have 
suggested that the turbulent viscosity is so large that the
entire interstellar medium becomes approximately spherical
(Nulsen, Stewart, \& Fabian 1984; Nulsen \& Fabian 1995),
but no detailed calculations were provided to support
this conjecture.

In the following section we describe rotating 
cooling flows in which 
angular momentum is lost by cooling dropout.  
Then we study the outward transport of 
angular momentum by turbulent viscosity and the associated 
diffusion of interstellar metallicity.

\section{COOLING DROPOUT IN ROTATING COOLING FLOWS}

In the earliest studies of non-rotating cooling flows, 
the interstellar gas was assumed to flow completely to 
the galactic center where, 
in the region of increasing gas density, 
it cooled by intense radiative losses. 
The notion that the hot interstellar gas in cooling flows 
cools not just at the centers of elliptical galaxies  but 
throughout the galactic volume was first advanced 
by Fabian, Nulsen \& Canizares (1982) and Thomas (1986). 
This distributed cooling was introduced to avoid the 
strong central peaking of the X-ray surface brightness $\Sigma_x$ 
that occurred in centrally-cooling 
models but not in the galaxies observed. 
Our recent studies 
indicate that the effect of volumetric cooling 
on $\Sigma_x$ is more complicated.
Although $\Sigma_x$ is globally reduced if the gas 
cools before it has flowed through the entire potential of the 
galaxy, this reduction is generally rather modest. 
However, in localized regions of intense cooling dropout 
$\Sigma_x$ actually {\it increases} because 
of the enhanced density and emission associated with  
cooling regions; if cooling dropout is restricted to a 
small but finite central region,  
$\Sigma_x$ can deviate even further from observed profiles. 

Notwithstanding these contrary results, the evidence for 
spatially distributed 
cooling of interstellar gas is apparent in other observations.
The dynamically determined masses of central black holes 
(Magorrian et al. 1998) are typically much less than 
the masses of interstellar gas that have cooled
over a Hubble time, typically
several $10^{10}$ $M_{\odot}$ 
in massive elliptical galaxies (Brighenti \& Mathews 1999b).
The cooled mass is also far larger than the masses 
of cold HI or H$_2$ gas observed in luminous elliptical galaxies 
(e.g., Bregman, Roberts \& Giovanelli 1988;
Braine, Henkel \& Wiklind 1997),
indirectly indicating that stars 
have formed in the cooled gas.
However, if the stars produced from cooled gas within $R_e$
are optically dark (e.g., brown dwarfs),
the thinness of the fundamental plane is difficult 
to understand (Mathews \& Brighenti 1999b). 
This suggests that the young, low-mass stars that form are 
optically luminous, which is also 
consistent with H$\beta$ indices observed in the 
stellar spectra in some luminous elliptical galaxies 
(Gonzalez 1993; 
Faber et al. 1995;
Trager 1997; Trager et al. 1998; 
Mathews \& Brighenti 1999c). 
Another indication of distributed interstellar cooling 
is the extended optical line emission observed 
in in the central regions of 
most or all bright elliptical galaxies 
(e.g., Macchetto et al. 1996).  
We have argued that this gas with temperature $\sim 10^4$ K 
is cooled interstellar gas ionized by stellar UV 
(Mathews \& Brighenti 1999a).

In the absence of 
spatially distributed mass dropout or 
angular momentum redistribution, 
all the cooled gas in rotating cooling flows is 
ultimately deposited in a large disk 
(Paper I; Brighenti \& Mathews 1997). 
Because the gas density and radiative emissivity increase 
as the cooling flow approaches the disk, the X-ray images 
in these cooling flow models are highly 
flattened unless the galaxy is viewed along the axis 
of rotation.
In the following discussion we show that 
spatially distributed mass dropout can sharply reduce,
but not eliminate, the rotational flattening of X-ray images. 
As a result, the approximate circular shapes 
of observed X-ray images may be the 
most convincing evidence for the existence of cooling dropout.
We begin by briefly discussing the gravitational potential 
and stellar dynamics in a simple rotating elliptical galaxy; 
this is followed with a description of 
a gas-dynamical model including 
the cooling dropout process.

\subsection{Model of a Rotating Galaxy}

For comparison with our earlier models of rotating elliptical 
galaxies and for mathematical simplicity, 
we adopt here the same E2 model galaxy described 
in detail in Paper I.
The ellipsoidal mass distributions for stars and dark 
matter in this galaxy are assumed 
to have King and pseudo-isothermal profiles respectively:
$$\rho_*(R,z) = \rho_{o*}
[1 + (R/R_{c*})^2 + (z/z_{c*})^2]^{-3/2}$$
$$\rho_h(R,z) = \rho_{oh}
[1 + (R/R_{ch})^2 + (z/z_{ch})^2]^{-1}.$$
Both distributions are truncated at the same large 
ellipsoidal boundary defined by 
$R_t$ and $z_t = R_t(1 - e^2)^{1/2}$. 
The eccentricity of an E2 galaxy ($n = 2$) is 
$e = [1 - (1 - n/10)^2 ]^{1/2} = 0.6$.
Scale parameters $R_{c*}$
and $R_{ch}$ for the 
E2 elliptical can be derived from a standard 
spherical E0 galaxy having the same central densities 
$\rho_{o*}$ and $\rho_{oh}$ by expanding the 
characteristic core radii $R_{c}^{(s)}$ 
for the spherical galaxy in the 
$R$-direction, $R_{c} = R_{c}^{(s)}(1 - 0.1n)^{-1/3}$.
The physical properties of the fiducial non-rotating 
E0 galaxy are listed in Table 1.
Values of $R_e$, $L_B$ and $\sigma_*$ have been chosen 
so that the E0 galaxy lies on the fundamental plane.  
The total mass of dark matter $M_{ht}$ is nine times 
that of the stellar component, $M_{*t}$.

As described in Paper I, we solve the two-dimensional 
Jeans equation to derive the velocity dispersion 
in the meridional plane, $\sigma^2(R,z) = \sigma_R^2 = \sigma_z^2$. 
Following Satoh (1980), we decompose the mean square azimuthal
speed into a random dispersion $\sigma_{\phi}^2$ and 
a systematic rotation $\overline{v_{\phi}}^2$ using a single 
parameter $k$ assumed to be constant,
$$\sigma_{\phi}^2 \equiv \overline{v_{\phi}~^2} -
\overline{v_{\phi}}^2 = k^2 \sigma^2
+( 1 - k^2) \overline{v_{\phi}~^2}.$$
When $k = 1$ the stellar velocity dispersion is 
isotropic and the ellipsoidal distortion 
is produced entirely by rotation; 
when $k = 0$ there is no net rotation and the 
ellipsoidal flattening 
is produced entirely by anisotropic velocity dispersion,
$\sigma_{\phi}^2 > \sigma^2$.
For a typical slowly rotating giant elliptical galaxy in which 
most (but not all) 
of the optical ellipticity is due to anisotropic orbits,
we choose $k = 0.5$ as in the ``E2;0.25'' model 
described in Paper I. 
From the Jeans equations we can determine 
a self consistent rotation 
for the stellar system, $v_{*\phi}(R,z)$, which is  
required for the gas dynamical equations.
The total galactic potential $\Phi(R,z)$, 
used in the gas dynamical equation of motion, 
can be derived 
from the density distributions above (see Paper I). 
The mean stellar temperature that appears 
in the gas dynamical equation for 
thermal energy is given by
$$T_*(R,z) = {\mu_m m_p \over 3 k_B} 
(2 \sigma^2 + \sigma_{\phi}^2 ),$$
where $\mu_m = 0.62$ is the mean molecular weight 
and $m_p$ is the proton mass.

\subsection{Cooling Dropout}

The physical origin of spatially localized interstellar cooling 
and mass dropout is a hypothetical ensemble of 
entropy fluctuations. 
For example, a small region in the 
cooling flow with low entropy 
(in pressure equilibrium with low $T$, high $\rho$) 
will radiate more efficiently and 
cool before the ambient flow reaches the galactic center 
or cooling disk. 
The entropy irregularities have complex origins -- supernova 
explosions, stellar mass loss, galactic winds 
and other disturbances in the early 
universe, inhomogeneous magnetic fields, etc. -- so it is not 
currently possible to derive them from first principles.
However, by exploring the influence of 
a variety of assumed mass dropout profiles on the X-ray 
surface brightness and radial mass distributions in luminous 
elliptical galaxies, 
Brighenti \& Mathews (1999b) confirmed that the following 
modification of the equation of continuity
gives adequate results:
$${\partial \rho \over \partial t}
+ {\bf \nabla}\cdot \rho {\bf v}
= \alpha_* \rho_* - q {\rho \over t_{do}}.$$

The negative term on the right hand side represents 
the disappearance of gas from the flow by cooling dropout.
This sink term is proportional to the local gas density
divided by the dropout time
$t_{do} = 5 m_p k T / 2 \mu_m \rho \Lambda$, 
the time for gas to
radiatively cool locally at constant pressure.
This kind of simple mass loss with $q = 1$, similar to that
used by Sarazin \& Ashe (1989), was shown by
Brighenti \& Mathews (1999b) to be consistent with optical
and X-ray observations of massive elliptical galaxies.

The first term on the right side of the equation 
of continuity above describes 
the source of new interstellar gas, which is proportional 
to the local stellar density and to  
$$\alpha_*(t) = \alpha_n [t/(t_n - t_{*s})]^{-1.3},$$
the specific mass loss from evolving stars. 
Here $\alpha_n = 4.7 \times 10^{-20}$ s$^{-1}$,  
$t_n = 13$ Gyrs is the current cosmic time and 
$t_{*s} = 1$ Gyr is the time that the galactic stars 
are assumed to have formed
(see Brighenti \& Mathews 1999a for details).
We are particularly interested in the flattening 
of X-ray images within $R_e$ from the centers of 
large ellipticals.
In this central region most of the interstellar gas 
is produced by stellar mass loss, so we shall 
generally ignore additional gas
flowing in from the galactic halo region beyond
the optical galaxy.

Interstellar gas is heated and enriched by Type Ia 
supernovae which we assume occur at a rate
SNu$(t) = $ SNu$(t_n)(t_n/t)$,
where SNu$(t_n) = 0.03$ SNIa
per 100 yrs per $10^{10}$ $L_{B\odot}$
(Cappellaro et al. 1997).
The specific rate of mass injection into the 
cooling flow by Type Ia supernovae is
$\alpha_{sn} = \nu_{sn} m_{sn}/ M_{*t}$
where $\nu_{sn} = {\rm SNu}(t) (L_B/10^{10})
/3.15 \times 10^9$ is the number of
supernovae per second. 
The rate that gas is supplied to the interstellar medium 
by normal stellar mass loss greatly exceeds the mass input 
from supernovae, i.e., $\alpha_* \gg \alpha_{sn}$.

\subsection{Rotating Cooling Flows with and without Dropout}

Figures 1a and 1b illustrate the X-ray surface brightness  
$\Sigma_x(R,z)$ in the ROSAT band (0.1 -- 2.4 keV) 
for the rotating elliptical at time $t_n = 13$
Gyrs when viewed along the equatorial plane 
and at 60$^o$ inclination.
As in Paper I, our 2D logarithmic 
computational grid extends beyond
100 kpc in both 
$R$ and $z$, but Figure 1 zooms in to 
show only the central $\sim$15 kpc region that is 
most relevant to published X-ray images.
Figure 1a shows the extremely flattened
X-ray appearance of the E2 galaxy when no distributed 
cooling dropout is present. 
This diagram is similar to Figure 12b of Paper I 
with several small differences due to the larger 
value of $t_n$ used in that paper and 
the slightly lower Type Ia supernova rate that we assume here.
When the same rotating cooling flow is observed at 
60$^o$ inclination, the bright 
X-ray emission associated with the outer disk 
is seen as a separate feature at $\sim 5$ kpc
(Figure 1b). 
The current ROSAT band luminosity for this galaxy within the 
optical image is $L_x = 2.0 \times 10^{40}$ ergs s$^{-1}$,
about 5 times less than the non-rotating version of the 
same galaxy (Paper I).
However, this $L_x$ must be regarded as 
an underestimate because our computational proceedure is 
inaccurate when most of the radiative 
cooling occurs in the central
zone (Brighenti \& Mathews 1999b).

Figures 1c and 1d show the same X-ray views of the rotating
galaxy at $t_n = 13$ Gyrs but now with cooling 
dropout ($q = 1$) and with all other 
parameters identical to those in Figures 1a and 1b.
The surface brightness 
$\Sigma_x(R,z)$ in Figures 1c and 1d includes the additional 
X-ray emission from the cooling regions, but the 
isophotal shapes 
are influenced very little by this contribution. 
It is immediately clear from Figure 1c that radiative dropout 
is a major explanation for the nearly circular appearance  
of observed X-ray images.
Evidently, interstellar gas is cooling from the flow 
before it has progressed
very far from its stellar origin toward the axis of rotation. 
However, the central X-ray bright core of 
the image is still noticeably flattened in Figure 1c 
with an X-ray ellipticity that is slightly 
greater than that of the E2 stellar image.
(The optical isophotal contours would intersect the 
$z$-axis at 0.8 of the $R$-axis intersection.) 
This excess ellipticity of the X-ray image is not 
greatly diminished if the galaxy is viewed at an inclination
$i = 60^o$ (Figure 1d). 
The ROSAT luminosity for the $q = 1$ model E2 galaxy 
at time 13 Gyrs is
$L_x = 8.8 \times 10^{40}$ ergs s$^{-1}$.

When there is no cooling dropout, $q = 0$ as in Figure 1a, 
a large and massive 
disk of cold gas forms in the equatorial plane, rotating at 
the local circular velocity; such disks are described in detail 
in Paper I and in Brighenti \& Mathews (1997).
However, in the calculation with 
$q = 1$ mass dropout by radiative losses 
as in Figure 1c, no disk of cold gas forms in 
the $z = 0$ plane at least 
beyond $\sim 150$ pc, the size of our 
innermost computational grid zone. 
This failure to form a cold disk 
results from the increased radiative dropout rate 
as the density in the cooling flow increases 
toward the equatorial plane,
$(\partial \rho / \partial t)_{do} \propto q \rho/t_{do}
\propto \rho^2$.

The azimuthal velocity distributions 
$v_{\phi}(R,z)$ in the interstellar gas 
shown in Figures 2a and 2b 
correspond to the cooling flows imaged 
in Figures 1a and 1c respectively; 
interstellar rotation is significantly reduced 
in the presence of cooling dropout.
Figure 2a shows the azimuthal velocity distribution 
at 13 Gyrs without radiative dropout.
As described in Paper I, without dropout mass loss 
there is no loss of angular momentum 
and the slow inward velocity of the cooling flow leads 
to a dramatic spinup in angular velocity. 
In Figure 2a the gas is flowing 
toward a cool disk that extends to $\sim 5$ kpc.
As interstellar gas converges toward the disk, 
its azimuthal velocity  
$v_{\phi}$ approaches the local circular velocity, 
$\sim 400$ km s$^{-1}$, as it cools and 
merges with the rotationally supported cooled disk.

With $q = 1$ (Figure 2b) the rotational velocities are reduced by 
approximately 30 percent in the region shown within 15 kpc 
of the center. 
If the entire major axis (-5 to +5 kpc) were observed, 
the rotational broadening of X-ray emission lines 
would be $\sim 400$ km s$^{-1}$, similar to the 
thermal width at the virial temperature, 
but resolution of such lines would be difficult or impossible 
even with the {\it Chandra} Observatory.
For purposes of comparison with future high-resolution X-ray 
observations, we show in Figure 2c the emission-weighted line of 
sight velocity at every position for the $q = 1$ calculation,
$$v_{los}(R,z) = {\int n_e n_p \Lambda_{ROSAT} 
{\mathbf v_{\phi}} \cdot {\mathbf n_s} ds  \over 
\int n_e n_p \Lambda_{ROSAT} ds}$$
where $\mathbf{n}_s$ is a unit vector along the line of 
sight represented with the $s$ coordinate. 
The contours in Figure 2b and 2c are rather similar 
due to the strong spatial gradient in electron density, 
but $v_{los}(R,z)$ is about 25 percent smaller 
than $v_{\phi}(R,z)$ over most of the region plotted.

Since it is likely that the cooled gas 
forms into optically luminous 
stars (Mathews \& Brighenti 1999b), it is of interest 
to compare in Figure 3 
the local density of cooled (dropout) 
gas $\rho_{do}(R,z)$
for the $q = 1$ solution
with that of the old stars, $\rho_*(R,z)$. 
Notice that the density distribution of 
cooled gas is considerably flatter than the E2 
distribution of the old stellar population; 
the isodensity ellipticity of the gas just prior to 
dropout reflects rotational flattening 
in the galactic potential.
The orbits of the young stellar population are expected
to inhabit the same flattened volume 
as the interstellar
dropout from which they formed, with  
stellar density contours similar to $\rho_{do}(R,z)$.
For an optically luminous population of young stars 
formed from cooled gas, 
we predict from Figure 3 
that the H$\beta$ equivalent width 
in the stellar spectrum 
(produced mostly by young stars) 
should exhibit a decidedly greater ellipticity 
than that of the bulk of the stellar light. 
In addition, since the interstellar gas 
in the cooling flow spins up 
before cooling, we expect that the young stellar population 
formed from mass dropout 
has a somewhat greater systemic rotation than that 
of the old stars.

For a larger dropout coefficient, $q = 4$, the X-ray 
images are even more circular and the corresponding 
azimuthal velocity $v_{\phi}(R,z)$ is lower than 
that of the $q = 1$ solution. 
However, when $q = 4$ the apparent gas temperature 
in the cooling flow is (unrealistically) lowered by the larger 
emission contribution from cooling sites.  
We have shown that the observed gas temperature profile in 
NGC 4472 cannot be fit if $q \ge 4$ 
(Brighenti \& Mathews 1999b).

\section{VISCOUS DIFFUSION OF ANGULAR MOMENTUM}

We now consider an alternative possible explanation 
of the circularity of X-ray images of elliptical galaxies: 
transport of angular momentum away from the 
axis of galactic rotation by turbulent diffusion.

Flow-induced turbulence can occur when the
ratio of inertial to viscous terms in the equation
of motion is large,
i.e. when the Reynolds number
${\cal R} = \rho u L / \mu$ is much larger than unity.
Here $\mu$ is the viscosity and $\rho u L$ represent
the characteristic density, velocity, and length scale
in the cooling flow.
In an unmagnetized plasma the ``molecular''
viscosity $\mu$ is
proportional to the plasma mean free path
$\lambda \propto T^2/n$ which can be very large in the 
hot interstellar gas. 
However, Faraday depolarization studies in the interstellar gas 
in elliptical galaxies (Garrington \& Conway 1991) indicate 
microgauss magnetic fields for which the 
proton Larmor radius $r_L$ is
much smaller than $\lambda$, greatly increasing
the effective Reynolds number and 
promoting the likelihood of turbulence.
To estimate the Reynolds number 
we replace $\lambda$ with 
the thermal gyroradius $r_L$ in the expression for 
the viscosity, 
$\mu \sim \rho v_{th} r_L$, where 
$v_{th} = (3 k_B T / m_p)^{1/2}$ is the thermal velocity.
The resulting Reynolds number for rotating cooling flows, 
$${\cal R} = \left({\rho u r \over \rho v_{th} r_L}\right)
\approx ~~~~~~~~~~~~~~~~~~~~~$$
$$
3 \times 10^{11} 
\left({v_{\phi} \over 200~{\rm km/s}}\right)
\left({r \over {\rm kpc}}\right) 
\left({T \over 10^7~{\rm K}}\right)^{-1/2}
\left({B \over \mu{\rm G}}\right),$$
is enormous, so turbulent flow can be expected.

The Rayleigh criterion establishes 
a necessary condition to avoid shear turbulence in rotating 
flows: for stability to linear perturbations the specific
angular momentum must increase from the rotation axis. 
By this criterion alone, rotating galactic 
cooling flows appear to be stable. 
Initially, 
as the gas first enters the flow, its specific angular momentum 
increases outward; this follows from the flat 
stellar rotation curves typically observed in slowly rotating
massive elliptical galaxies (e.g. van der Marel 1991).
We have verified that the sense of this 
gradient in the angular momentum density 
is preserved in the final 
cooling flows shown in Figure 1a.

However, rotating flows that satisfy 
the Rayleigh criterion may still be unstable to non-infinitesimal
perturbations that result from turbulent mixing. 
A variety of non-linear perturbations 
are present in elliptical galaxies which can generate
interstellar turbulence: transport and dissipation of 
mass ejected from orbiting stars, Type Ia supernovae,
the radial sinking of denser cooling sites, 
buoyant and shearing magnetic fields, etc.
Indeed, interstellar mixing 
(turbulence) due to stellar and supernova sources 
is likely to generate large magnetic fields 
in the cooling flow gas 
by the turbulent dynamo process.
These fields, which can serve to redistribute 
angular momentum, 
may become dynamically important in the central 
regions of the flow where the magnetic energy density 
is further increased by the converging flow 
(Soker \& Sarazin 1990;
Moss \& Shukurov 1996; Mathews \& Brighenti 1997).
However, if interstellar turbulence is invoked to account 
for the circularization of X-ray images, the associated 
spatial diffusion must not 
suppress abundance gradients observed in the hot interstellar 
gas (Matsushita 1997).

In the following discussion 
we estimate the effective viscosity corresponding 
to various sources of interstellar mixing or turbulence.
Then we describe modifications to the gas dynamical equations
that incorporate a spatially dependent turbulent viscosity. 
This is followed by a description of rotating cooling flows 
including both mass dropout and turbulent viscosity.

\subsection{Sources of Interstellar Viscosity}

The effective viscosity due to stellar mass loss and 
Type Ia supernovae can be estimated 
by constructing the viscosity 
from its simplest kinematical representation, 
$\mu \approx \rho \ell v $, where $\rho$ is the local gas density,
$\ell$ is the mean free path or coherence length 
of the disturbed region, and $v$ is the 
characteristic velocity appropriate to each process.
To estimate the viscosity 
$\mu_*$ generated by stellar mass loss,
we use a length scale 
$\ell_* \approx 2a_{en} = 4.5 \times 10^9 \rho^{-1/3}$ 
based on the radius $a_{en}$ of a recently ejected stellar 
envelope in pressure equilibrium 
(Mathews 1990) and the mean turbulent 
velocity $v_{*t} = 5.7 \times 10^5 r_{kpc}^{-0.2}$ cm s$^{-1}$ 
as estimated by Mathews and Brighenti (1997).
In evaluating $v_{*t}$ we use a stellar dispersion 
$\sigma_* = 351$ km s$^{-1}$ (Table 1) and assume that  
half of the kinetic energy in stellar ejecta 
supports the turbulence.
With these dimensional values and taking 
$\rho \approx 1.05 \times 10^{-25} r_{kpc}^{-1.8}$ gm 
cm$^{-3}$ we find
$\mu_* \approx 0.06 r_{kpc}^{-1.4}$ gm cm$^{-1}$ s$^{-1}$ for 
the effective viscosity generated by stellar ejecta.
Alternatively, the characteristic length $\ell_*$ could 
be based on the total volume of interstellar 
gas disturbed by the motion of an ejected stellar envelope.
If the stopping distance of the stellar envelope 
after ejection from an orbiting star is 
$\ell_{st} \approx 1.2 \times 10^{22} r_{kpc}^{0.6}$
cm (Mathews 1990), then 
$\ell_* \approx \ell_{vol} \equiv  
(\pi a_{en}^2 \ell_{st})^{1/3}\
\approx 3.3 \times 10^{19}~r_{kpc}^{0.6}$ cm.
In this case the viscosity is somewhat larger, 
$\mu_* \approx \rho v_{*t} \ell_{vol} \approx
1.0 ~r_{kpc}^{0.6}$ gm cm$^{-1}$ s$^{-1}$, 
reflecting the uncertainty in estimating $\mu_*$.

For an equally approximate 
estimate of the viscosity corresponding to 
supernova remnants, we use an eddy size 
$\ell_{sn} = 3.6 \times 10^{11} \rho^{1/3}$ cm 
based on the blast wave calculations of Mathews (1990)
and a characteristic mixing 
velocity $v_{sn} = 2.0 \times 10^6 r_{kpc}^{-0.2}$ 
cm s$^{-1}$ from Mathews \& Brighenti (1997), 
assuming that half of the supernova energy goes into 
interstellar turbulence. 
The resulting viscosity for Type Ia supernova explosions is 
$\mu_{sn} \approx 16.2 r_{kpc}^{-1.4}$ gm cm$^{-1}$ s$^{-1}$. 
For both 
$\mu_*$ and $\mu_{sn}$ we ignore a weak time dependence 
due to variable mass loss from evolving stars 
and the time dependent Type Ia supernova rate respectively.
However, both $\mu_*$ and $\mu_{sn}$ decline 
with galactic radius in a similar fashion 
due to the decrease in gas and stellar density, so  
$\mu_{sn} \gg \mu_*$ is expected at all $r_{kpc}$ from 
these formal estimates.

An additional viscosity arises from the cooling dropout 
process. 
The increased density in a cooling site relative to the 
surrounding cooling flow 
causes it to sink more rapidly in the galactic potential 
than the background flow. 
The net effect of the ensemble of differentially 
sinking sites is an approximately radial exchange 
of gas in the flow with a corresponding 
viscosity.
When a site of density $\rho_s > \rho$ 
falls at its terminal velocity $u_t$, 
the drag force on the moving site 
must balance the gravitational force, 
$\rho u_t^2 \pi r_s^2 \approx (4 \pi / 3) r_s^3 \rho_s g$, 
where $g \approx 5.2 \times 10^{-7} r_{kpc}^{-1}$ cm s$^{-2}$ 
is an approximation to the galactic gravitational acceleration.
The radius $r_s$ of a cooling site can be estimated 
using the approximate model for an isobaric, steady state
cooling site described
by Mathews \& Brighenti (1999a). 
Within the cooling site structure, the temperature
varies nearly linearly with radius, 
$T = [K \Lambda(T)]^{1/3} r$ where
$K \Lambda = (8 \pi/5) (\mu_m/k)^3 P^2 m_p f \Lambda(T)
/{\dot m}_s = 5.2 \times 10^{-38} r_{kpc}^{-3.6}$.
Here $\mu_m = 0.62$ is the molecular weight, 
$f(\mu_m) = 0.58$,
and $P \approx 1.9 \times 10^{-10} r_{kpc}^{-1.8}$ 
dyne is the
approximate interstellar pressure variation.
The flow into each site is ${\dot m}_s = 10^{-6}$
$M_{\odot}$ yr$^{-1}$ since line emission from 
$\sim 10^6$ sites is required to match the
total H$\beta$ luminosity from luminous elliptical 
galaxies 
(Mathews \& Brighenti 1999a) and approximately 1 $M_{\odot}$
of interstellar gas cools each year. 
Taking $T \approx 0$ at the site center, the radius where 
$T = 10^6$ K, at which the site density is $10 \rho$, 
is $r_s \approx 2.7 \times 10^{18} r_{kpc}^{1.2}$ cm 
and the corresponding terminal velocity is 
$u_t = [4 g r_s (\rho_s/\rho)/3]^{1/2} 
= 43~r_{kpc}~^{0.1}$ km s$^{-1}$.
The effective viscosity is then 
$\mu_s \approx n_s m_s u_t r_s = q \rho u_t r_s$
where the product of the space density of sites $n_s$ and 
the local site accretion rate 
${\dot m}_s$ is set equal to $q \rho /t_{do}$ 
and we assume $m_s/{\dot m}_s = t_{do}$.
Combining all these results, the effective viscosity for 
the ensemble of sinking cooling sites is roughly 
$\mu_s \approx 1.2~r_{kpc}^{-0.5}$ gm cm$^{-1}$ s$^{-1}$. 
This viscosity is small within most of the bright X-ray 
core of the galaxy. 
Although $\mu_s$ exceeds $\mu_{sn}$ for $r \gta 50$ kpc, 
there is little or no optical line emission from these 
distant regions.

Obviously, each of these estimates for $\mu_*$, $\mu_{sn}$ and 
$\mu_s$ is highly uncertain, 
perhaps good to an order of magnitude.  
In addition to these possible sources of turbulent viscosity, 
the interstellar gas 
could become turbulent due to continuous mergers
of gas-rich smaller galaxies having turbulent wakes, 
but there is little optical evidence 
for sufficiently numerous merging galaxies. 
The torquing action of magnetic stresses can 
also propagate angular 
momentum outward, masquerading as viscosity. 
Finally, we note the considerable energy available in 
differential shear that could be converted to turbulence 
if the rotating flow is globally unstable.
If the characteristic change of azimuthal 
velocity over length $\ell_{\phi}$ is 
$\Delta v_{\phi} \approx 
| \partial v_{\phi} / \partial R| \ell_{\phi}$, then 
the effective viscosity is
$\mu_{\phi} \sim \rho | \partial v_{\phi} / \partial R| \ell_{\phi}^2$.
From Figure 2a, $| \partial v_{\phi} / \partial R| \approx 
40$ km s$^{-1}$ kpc$^{-1}$ so 
$\mu_{\phi}  \sim 1000 r_{kpc}^{-1.8} \ell_{\phi,kpc}^2$, 
which could be large depending on the unknown scale length
$\ell_{\phi}$.

\subsection{Viscous Terms}

In rotating cooling flows the flow velocity in the meridional 
plane is very subsonic, typically more than  
an order of magnitude smaller than the azimuthal velocity 
(Paper I).
Therefore, only 
the azimuthal velocity component $v_{\phi}$
is relevant to our investigation of the importance of 
the viscous transport of angular momentum 
in circularizing the X-ray images 
of elliptical galaxies.
In our numerical solutions, we implement 
the contribution of viscosity 
by solving the conservation equation for 
angular momentum density $\rho h \equiv \rho R v_{\phi}$. 
The rate of change of angular momentum density due to viscous 
effects alone is given by 
$$\left({\partial ( \rho h) \over \partial t}\right)_{vis} = 
\mu {\partial^2 h \over \partial R^2}
+ \mu {\partial^2 h \over \partial z^2}
-{\mu \over R} {\partial h \over \partial R}~~~~~~~~~~~$$
$$~~~~~~~~~~~~-{2 h \over R} {\partial \mu \over \partial R}
+{\partial h \over \partial R}{\partial \mu \over \partial R}
+{\partial h \over \partial z}{\partial \mu \over \partial z}$$
where $\mu = \mu(R,z)$ is a spatially 
variable viscosity.
We have incorporated these viscous terms 
into the Eulerian ZEUS2D code (Stone \& Norman 1992)
as a separate computational process. 
In addition to the usual consideration of ``source'' and 
``transport'' steps in the Eulerian numerical procedure, 
we modify $v_{\phi}$ for viscous transport 
by solving the equation above using a  
standard operator-splitting procedure.
The viscosity terms in the angular momentum equation above
correspond to a modified two-dimensional diffusion.
We solve these terms 
at each timestep using a second order 
alternating-direction implicit (ADI) difference scheme
(Press et al. 1992) on the 2D logarithmic grid. 
Angular momentum diffuses spatially because 
of viscous effects. 
When there is no cooling dropout,
our numerical procedure conserves 
total angular momentum but also allows it to flow 
out of the computational grid at the boundaries.

The equation for thermal energy conservation must 
also be corrected for the contribution 
of viscous dissipation.
An additional numerical step is applied to the 
thermal energy equation 
to solve for the following dissipative heating term:
$$ \left(\rho {d \varepsilon \over dt}\right)_{vis} 
= \mu R^2 \left\{ 
\left[ {\partial  \over \partial R}
\left( {h \over R^2} \right) \right]^2
+ \left[ {\partial  \over \partial z}
\left( {h \over R^2}\right) \right]^2 
\right\}$$
where $\varepsilon = 3 k T/ 2 \mu_m m_p$ 
is the specific thermal energy.
The remaining (non-viscous) 
terms in the thermal energy equation, and the
complete equation of motion, are discussed 
in Brighenti \& Mathews (1999a).

\subsection{Effect of Viscosity on X-ray Images}

In general we find that the circularization of 
X-ray images of elliptical galaxies is easier to 
achieve with cooling dropout than with viscous 
redistribution of angular momentum. 
For this reason we begin our discussion of 
viscous rotating cooling 
flows with the supernova viscosity, 
the largest viscosity expected from the estimates above.
In this case $\mu = \mu_{sn}(r)$ where 
$r = [R^2 + z^2]^{1/2}$ is the radial distance
from the galactic center.
Figure 4a shows the X-ray image of our rotating galaxy 
viewed along the equatorial plane 
at time $t_n = 13$ Gyrs but now including viscous terms 
based on $\mu_{sn} = 16.2~r_{kpc}^{-1.4}$ gm cm$^{-1}$ s$^{-1}$. 
The solution shown in Figure 4a also includes 
a mass dropout coefficient $q = 1$ 
and the additional surface brightness $\Sigma_x(R,z)$ 
contribution from the collective 
emission from cooling sites. 
The contours in Figure 4a are 
seen to be almost identical 
to those shown in Figure 1c, the same 
calculation without viscosity. 
The velocity field (not shown) is also nearly the same 
as that in Figure 2b 
and the spatial distribution of dropout gas in this
solution $\rho_{do}(R,z)$
is almost identical to the same non-viscous
$q = 1$ solution shown in Figure 3.
It is clear that the additional small viscosity 
due to stellar mass loss 
and moving cooling sites, $\mu_*$ and $\mu_s$, 
will have almost no effect on the solutions.
We conclude that the expected level of turbulent viscosity 
is insufficient to circularize the X-ray images 
of cooling flows in elliptical galaxies.

However, our estimates of the viscosity are extremely 
crude, and it is possible that the real value is 
considerably larger. 
To explore this idea, we show in Figure 4b 
surface brightness  
$\Sigma_x(R,z)$ contours at $t_n = 13$ Gyrs 
computed with a greatly 
enhanced viscosity, $\mu = 100 r_{kpc}^{-1.4}$, 
which still has the decreasing radial dependence 
expected if the viscosity has a stellar origin.
The surface brightness contours in Figure 4b,
also computed with $q = 1$ and dropout emission,
are noticeably rounder in the bright central region.
In addition, a projection of the X-ray emission from 
this solution viewed at inclination $i = 60^o$ 
(Figure 4c)
has contours that are considerably more 
circular and compares favorably with the 
observations of Hanlan \& Bregman (1999). 

Finally, we consider
solutions with very large and uniform turbulent viscosity. 
Our estimates of the sources of viscosity 
-- $\mu_{sn}$, $\mu_*$, and $\mu_s$ -- are independent 
of galactic shear. 
However, as we speculated earlier, 
it is possible that these
(or other) non-linear perturbations could unleash the 
considerable energy density in the shear flow to maintain
a high level of turbulence throughout the flow.
Strong, approximately uniform distributed turbulence
might result from continuing mergers of small gas-rich
galaxies penetrating through the cooling flow.
But this hypothesis
seems doubtful since there is little or no evidence
for this type of continuous galactic bombardment
in either X-ray or optical images.
If many on-going small mergers were required
to establish global interstellar turbulence,
numerous small, gas-rich galaxies should be
apparent in optical images of large elliptical
galaxies within the optical image and beyond.

Another obvious 
motivation for exploring larger turbulent viscosities is the 
persistence of X-ray flattening in the 10 - 15 kpc 
region apparent in Figures 1b, 1d, 4b and 4c, 
which may still exceed X-ray ellipticities in 
{\it Einstein} HRI and ROSAT HRI images.
Isophotal flattening (and twists) have been reported
by Boute \& Canizares (1998) in some large ellipticals,
but this effect occurs at much larger radii and 
is probably related to misalignments of the dark 
halo and stellar potentials which we do not consider here.

In Figure 5 we show equator-on views of the 
X-ray surface brightness at 13 Gyrs  
for solutions computed 
with large and 
spatially uniform viscosities, $\mu = 5$ and 20 
gm cm$^{-1}$ s$^{-1}$ respectively.
Both solutions include dropout with $q = 1$ and
emission produced by cooling sites.
Clearly, the largest turbulent viscosity 
($\mu = 20$) can 
circularize images both in the 10 - 15 kpc region 
and near the bright center. 
Turbulent viscosity has very little influence on the 
current ROSAT luminosity within the optical image of the 
E2 galaxy: 
$L_x = 9.1$, 11.0, 10.0, and $11.5 \times 10^{40}$ 
ergs s$^{-1}$ for $q = 1$ models with 
the four viscosities we consider,
$\mu = 16.2 r_{kpc}^{-1.4}$, $100r_{kpc}^{-1.4}$, 
5, and 10 respectively.

As we mentioned earlier, the fraction of
interstellar gas that derives from stellar mass
loss decreases with galactic radius. 
At $r \gta 10$ kpc a larger fraction of the
gas has flowed in from the halo 
(Brighenti \& Mathews 1999a) and is not created
solely by stellar mass loss as we have assumed. 
The angular momentum density in gas flowing in from the
outer halo may be different
(and possibly larger) than that of the stars.
However, if most of this extended gas was expelled 
from the large elliptical at early times, 
driven by a Type II supernova wind,
its angular momentum may differ little from 
that of the galactic stars.
But, even in this scenario, 
if this extended gas has been tidally influenced by 
other nearby galaxies, as may 
be common, when it flows in from the outer halo 
its specific angular momentum 
could differ from the galactic stars both 
in magnitude and direction. 

To explore the possible influence of
inflowing circumgalactic gas on the apparent X-ray
shape of galactic cooling flows, we performed 
several idealized calculations in which additional, 
non-rotating circumgalactic 
gas flows in from the galactic halo. 
In these models (not illustrated here)  
we used the same parameters for the circumgalactic 
gas as described in Brighenti \& Mathews (1998).
As expected, the ellipticity of the 
X-ray isophotes at 
10 - 15 kpc is reduced by the inflowing gas, 
but the effect is not very large. 
Evidently the small but significant global
spinup resulting from interaction with gas
expelled from the rotating stellar system 
is important even at these radii.

\section{VISCOUS DIFFUSION OF METALLICITY}

If viscosity is invoked to circularize the X-ray images 
of elliptical galaxies or to amplify magnetic fields in 
the interstellar gas (Mathews \& Brighenti 1997), 
then the level of turbulence required
must be consistent with 
observed metallicity gradients in the interstellar gas.
When X-ray data reduction procedures are applied 
to spatially resolved elliptical galaxies,
the iron abundance in the interstellar gas has been 
found to decrease with increasing radius approximately 
as $z_{Fe} \propto r^{-0.5}$ (e.g., Matsushita 1997). 
We restrict our discussion here to iron for which 
current data is most plentiful.
In the presence of spatial 
diffusion the continuity equation for iron mass is 
\begin{equation}
{\partial \rho c \over \partial t}
+ {\mathbf \nabla} {\mathbf \cdot} (\rho c {\mathbf u})
= {\mathbf \nabla \cdot} (\rho D {\mathbf \nabla} c) 
~~~~~~~~~~~~~~~~~~~~~~~~~~
\end{equation}
$$~~~~~~~~~+ \left[ 
\left({\alpha_* z_* \over 1.4} \right) 
+ \left( {\alpha_{sn}~ y_{Fe,Ia} \over m_{sn}}\right)
\right] \rho_*
- q {\rho \over t_{do}}c,$$
where $c \equiv c_{Fe} = \rho_{Fe}/\rho$ is the concentration 
of iron relative to the total mass density.
The abundance of iron in the hot gas 
relative to hydrogen mass can be found from 
the concentration by correcting 
for the mass of helium, $z = 1.4c$.
The diffusion coefficient depends 
on a characteristic velocity and mixing length, 
$D \approx v_d \ell_d $ and can be expressed 
directly in terms of 
the viscosity, $D \approx \mu / \rho$. 
Iron enters the interstellar gas both from normal 
stellar mass loss and from Type Ia supernovae.
For the stellar iron abundance we assume 
$$z_*(R,z) = z_{Fe,o} 
[1 + (R/R_{c*})^2 + (z/z_{c*})^2]^{-0.3}$$
with central value $z_{Fe,o} = 1.77 \times 10^{-3}$ 
(e.g., Arimoto et al. 1997;
Ishimaru \& Arimoto 1997).
Each supernova expels $y_{Fe,Ia} = 0.744$ $M_{\odot}$ 
in iron and a total mass of $m_{sn} = 1.4$ $M_{\odot}$. 
With a stellar mass to light ratio 
$M_{*t}/L_B = 9.14$,
we find $\alpha_{sn} = 4.85 \times 10^{-21}
{\rm SNu}(t_n) (t/t_n)^{-1}$ s$^{-1}$.

For most cases of interest 
viscous spatial diffusion is insufficient 
to flatten iron abundance gradients since the 
gradient is 
re-established on short timescales by Type Ia 
supernovae. 
The timescale for spatial diffusion of iron is 
$t_{d,Fe} \approx \rho r^2/\mu \approx 2 \times 10^9 r_{kpc}^{1.6}$ 
years using $\rho \approx 1.05 \times 10^{-25} r_{kpc}^{-1.8}$ gm
cm$^{-3}$ and $\mu  = \mu_{sn} = 16.2 r_{kpc}^{-1.4}$ 
gm cm$^{-1}$ s$^{-1}$. 
By comparison, the timescale required 
for Type Ia supernovae to 
restore an interstellar iron concentration of 
$c = z/1.4$ is 
$t_{sn,Fe} \approx \rho c m_{sn}/ \alpha_{sn} y_{Fe,Ia} \rho_*$. 
Adopting $z = 0.5$, $z_{Fe,\odot} = 1.77 \times 10^{-3}$,
$\rho = 1.05 \times 10^{-25} r_{kpc}^{-1.8}$ gm cm$^{-3}$,
$\rho_* = 4.35 \times 10^{-22} r_{kpc}^{-3}$ gm cm$^{-3}$, 
and SNu$(t_n) = 0.03$, we find
$t_{sn,Fe} \approx 6 \times 10^7 r_{kpc}^{1.2}$ years. 
Therefore $t_d \gg t_{sn,Fe}$ and supernovae 
can maintain interstellar 
iron abundance gradients even in the presence 
of supernova-driven turbulence. 

To test this more quantitatively, 
we have solved the iron concentration equation 
in several evolving cooling flows.
In cylindrical geometry the divergence term 
in the iron continuity equation (1) is 
$${\mathbf \nabla \cdot} (\rho D {\mathbf \nabla} c) 
= {1 \over R} {\partial \over \partial R}
\left( R \rho D {\partial c \over \partial R} \right)
+ {\partial \over \partial z} 
\left( \rho D {\partial c \over \partial z} \right).$$
This diffusion equation was solved using a
second-order ADI 
method similar to that used for the viscosity terms in 
the equation of motion.

Figure 6 compares iron abundance profiles 
in two cooling flows at $t = t_n$ along 
the $R$ and $z$ axes for both the gas and stars. 
For the flow with $q = 1$, 
$\mu  = \mu_{sn} = 16.2 r_{kpc}^{-1.4}$
gm cm$^{-1}$ s$^{-1}$ and $D = \mu/\rho$, 
the diffusion has almost no effect on the 
iron distribution in the hot gas, in agreement 
with our simple estimate above. 
Even when the viscosity is increased 
to $\mu = 100$ throughout the 
computational grid, also shown in Figure 6, an iron 
abundance gradient is still present, but 
is beginning to flatten, particularly at 
the outer edge of the galactic stellar distribution
at $\log R \sim \log z \sim 2$.
This indicates that observed interstellar iron 
abundance gradients can be maintained only if 
the average turbulent viscosity is $\mu \lta 100$
gm cm$^{-1}$ s$^{-1}$.
To investigate further the importance of 
the source terms on the metallicity gradient,
we performed an 
additional exploratory 
computation (with $q = 1$ and $\mu = 100$) 
in which the iron source terms in equation (1) 
were set to zero beginning at time $t = 10$ Gyrs.
By time $t = 11$ Gyrs in this calculation 
the iron gradient in the interstellar 
gas was completely 
leveled, resulting in uniform $z_{Fe}(R,z)$ 
throughout the computational grid.
In this unrealistic test calculation our solution 
of the concentration 
diffusion term above behaved as expected.

The diffusive 
mixing of supernova enrichment in the interstellar 
gas is relevant to the possibility that 
locally metal-rich regions may cool by X-radiation and 
dropout faster than 
the rest of the cooling flow.
We estimate the metallicity diffusion time by 
equating the diffusion length from each 
supernova blast wave bubble to the mean 
distance between bubbles.
The number of Type Ia supernova bubbles in a large 
elliptical galaxy at any time,
${\cal N}_{bubb} \approx \nu_{sn} t_{bubb}
\sim 1.5 \times 10^3 (t_{bubb}/10^6~{\rm yrs})$, 
is quite small. 
Here $t_{bubb} \approx 10^6$ yrs is a typical 
lifetime of a supernova bubble to destruction 
by buoyant mixing (Mathews 1990).
However, the iron deposited in the cooling flow 
by Type Ia supernova persists long after the 
thermal energy in the hot 
blast wave bubble has dissipated.
To estimate the efficiency that supernova iron
mixes throughout the interstellar gas 
by turbulent diffusion, 
we imagine a time $t = 0$ at which the supernova 
explosions begin.  
Then the mixing time $t_{mix}$ for supernova iron 
to diffuse between supernova sites 
is found by equating the decreasing distance between 
supernova bubbles at $t > 0$ to the distance that 
iron has diffused by turbulent mixing from any 
particular site; $t_{mix}$ must then 
be compared to the local cooling flow time.
As Type Ia supernovae events increase with time $t$, 
the mean separation between sites is 
$\delta_{bubb} \sim (t 
\nu_{sn} \rho_* / M_{*t})^{-1/3} 
\sim 36 r_{kpc} t_{yr}^{-1/3}$ kpc,
estimated with 
$\rho_* = 4.35 \times 10^{-22} r_{kpc}^{-3}$ gm cm$^{-3}$.
This can be compared with 
the distance $\delta_{diff}$ that diffusion 
mixes metals from an individual supernova bubble in 
the same time $t$: 
$\delta_{diff} =( \mu t / \rho)^{1/2}
\sim 2.3 \times 10^{-5} t_{yr}^{0.5} r_{kpc}^{0.9}$ kpc
where we assume $\mu = \mu_{sn} = 
16.2 r_{kpc}^{-1.4}$ gm cm$^{-1}$ s$^{-1}$ and
$\rho = 1.05 \times 10^{-25} r_{kpc}^{-1.8}$ gm cm$^{-3}$.
Diffusive mixing of supernova 
enrichment in the interstellar medium 
occurs in time $t = t_{mix}$ found by setting 
$\delta_{diff} \approx \delta_{bubb}$, 
which gives
$t_{mix} \sim 3 \times 10^7 r_{kpc}^{0.96}$ yrs.
This mixing time is comparable to the 
radial flow time in the cooling flow,
$t_{flow} \sim r/u \sim 5 \times 10^7 r_{kpc} 
[u/20({\rm km}/{\rm s})]^{-1}$ yrs.
We conclude that supernova 
enrichment products are well mixed throughout 
the cooling flow for the Type Ia supernova 
rate we assume here; for turbulent viscosities 
$\mu \gta \mu_{sn}$, mixing is guaranteed.
In a cooling flow 
the local radiative cooling time is comparable 
to the flow time, so iron produced by supernovae 
disappears from the flow by mass dropout 
in time $\sim t_{flow}$.
The complete absence of diffusive mixing, 
as envisioned by Fujita, Fukumoto \& Okoshi (1996; 1997), 
in which unmixed, metal-rich supernova bubbles 
and stellar ejecta 
cool rapidly by enhanced X-ray line emission  
and differentially drop out from the flow 
may not be realistic, but we have not explicitly 
ruled out this possiblilty.

\section{SUMMARY AND CONCLUSIONS}

In this paper we have addressed the curious absence of 
X-ray disks in rotating, luminous elliptical galaxies.
In our previous hydrodynamic studies of rotating cooling 
flows (Paper I; Brighenti \& Mathews 1997) 
both angular momentum and mass were conserved 
and the resulting X-ray images were dramatically flattened 
out to 1-2 effective radii when 
viewed perpendicular to the axis of rotation.
These results appear to be inconsistent 
with the more circular X-ray images found by 
observations with {\it Einstein} HRI
(Fabbiano, Kim \& Trinchieri 1992) 
and ROSAT HRI (Hanlan \& Bregman 1999). 
Nevertheless, the X-ray shapes of 
elliptical galaxies are difficult to determine 
from the observations 
and we may need to wait for the next generation of 
X-ray telescopes with higher spatial resolution for 
definitive determinations of X-ray ellipticities. 
It is likely that interstellar rotational flattening will 
eventually be observed.

In the preceding discussion we have explored the 
circularizing influence of 
mass dropout and turbulent transfer of angular momentum
on the X-ray images of rotating galactic cooling flows.
Strong theoretical and observational arguments support 
both of these possibilities.
Localized radiative cooling leading to 
mass dropout throughout the cooling flow 
is indicated by the relatively low masses
of central black holes and adjacent nuclear regions 
when compared to the mass of interstellar gas that 
is expected to cool over cosmic time. 
Further evidence for cooling dropout is 
provided by optical line emission within $R_e$ and 
the evidence in this same region 
for a young stellar population 
having masses that extend up to 1 - 2 $M_{\odot}$ 
but not beyond (Ferland, Fabian, \& Johnstone 1994; 
Mathews \& Brighenti 1999a).
Interstellar turbulence is also very likely in elliptical 
galaxies, generated by mass transport associated with 
stellar mass loss, Type Ia supernovae and mass dropout. 
Indirect evidence for cooling flow 
turbulence is provided by interstellar 
magnetic fields of several $\mu$G at $\sim 10 R_e$,  
which are typically observed in elliptical galaxies having double 
radio sources.
Fields of this magnitude can be understood as originating from 
small 
seed fields ejected from mass-losing stars, followed by 
subsequent field amplification by 
an interstellar turbulent dynamo mechanism (Moss \& Shukurov 1996;
Mathews \& Brighenti 1997). 
Due to the field concentration associated with the 
inward, converging motion of the cooling flow, 
interstellar magnetic fields are further magnified and 
are expected to be particularly strong, perhaps exceeding 
equipartition values, in the central regions of cooling flows.
Owen and Eilek (1998) find 
fields of 10 - 100 $\mu$G within  
$r = 50$ kpc in the bright cD galaxy NGC 6166. 

Elliptical galaxies of lower optical 
luminosity $L_B \lta 3 \times 10^{10}$ 
$L_{B,\odot}$ differ in many qualitative ways from the 
more massive elliptical galaxies similar 
to the one we have studied here (Faber et al. 1997). 
Low luminosity elliptical galaxies 
rotate faster and are rotationally flattened.
We have shown (Brighenti \& Mathews 1997) that rotating, 
non-dropout cooling flows in these galaxies 
also form massive, extended disks of cooled gas.
However, when $q = 1$ mass dropout is included in these 
calculations, disks of cold HI (or H$_2$) do not form 
from the rotating cooling flow. 
There is observational evidence for extended disks
of cold HI gas in some low luminosity elliptical galaxies, 
but these HI disks often extend far beyond the optical 
images of the galaxies, indicating that they were created 
by a (possibly very old) merging event rather 
than by cooling flow dropout
(Oosterloo, Morganti \& Sadler 1999a; 1999b).
In some cases the cold gaseous disks have central holes
in neutral hydrogen replaced with HII emission 
(Oosterloo, Morganti \& Sadler 1999),
suggesting a more complete conversion into stars in this
part of the disk where gas pressures and densities
are highest and star formation (and ionization by 
PAGB stellar UV) should be most efficient.
In general,
systematically larger H$\beta$ features are observed 
in the stellar spectra of elliptical galaxies with 
$L_B \lta 3 \times 10^{10}$
$L_{B,\odot}$, 
indicating considerable 
recent star formation (de Jong \& Davies 1997). 
Unfortunately, 
thermal X-ray emission from the hot interstellar gas in these
low-luminosity, low-$\Sigma_x$ ellipticals is masked by 
the collective emission from low mass X-ray binaries. 
A detection of X-ray 
rotational flattening in their cooling 
flows will require spectral separation of the hot gas 
and stellar X-rays with high spatial resolution, 
clearly a job for the next generation of X-ray 
telescopes.

When mass dropout and turbulence are considered in 
rotating cooling flows inside luminous elliptical 
galaxies, we find that: 

\begin{itemize}
\item[(1)]{
Mass dropout alone strongly circularizes the 
X-ray images in rotating cooling flows since gas is 
removed from the flow before it has moved very far from its 
point of origin (stellar mass loss) toward the axis of rotation.
}
\item[(2)]{
Conversely, 
the absence of strong rotational flattening in X-ray 
images of elliptical galaxies is persuasive evidence 
for distributed mass dropout.
}
\item[(3)]{
In the presence of mass dropout by radiative cooling 
with $q = 1$, no cold gaseous disks form on the equatorial 
plane having radii 
larger than our innermost grid size, $\sim 150$ pc. 
This is due to the sensitivity of 
radiative cooling (and associated mass dropout) 
to the local density,
$(\partial \rho / \partial t)_{do} \propto q \rho/t_{do} 
\propto \rho^2$.
}
\item[(4)]{The spatial 
distribution of cooled gas mass is markedly 
flatter than that of the old stellar population, 
especially near the galactic center.
Any optical signature of a younger stellar population 
formed from the cooled gas, such as the H$\beta$ index, 
should also exhibit a significantly 
higher ellipticity than that of the older, background stars.
}
\item[(5)]{
The estimated viscosity from known sources of interstellar 
turbulence is dominated by motions induced by 
Type Ia supernovae; stellar mass loss and mass dropout 
are smaller sources of interstellar turbulent viscosity.
}
\item[(6)]{
Supernova-induced turbulent viscosity 
is insufficient to circularize the X-ray appearance of 
rotating 
elliptical galaxies; the viscous effects of 
stellar mass loss and cooling dropout 
on the X-ray images are even smaller.
}
\item[(7)]{
Much larger, spatially uniform turbulent viscosities 
can circularize the X-ray isophotes 
throughout the galaxy;  
such turbulent viscosities could result from 
rotational shear instabilities, 
but we have not demonstrated this instability here. 
}
\item[(8)]{
The spatial 
diffusion of interstellar iron due to turbulent 
viscosity does not appreciably reduce the interstellar 
iron abundance gradients observed in bright elliptical 
galaxies, provided the viscosity does not 
exceed $\sim 100$ gm cm$^{-1}$ s$^{-1}$. 
This follows since the 
observed interstellar iron 
gradients are re-established on 
short time scales by Type Ia supernovae and, to a 
lesser extent, by stellar mass loss.
}
\end{itemize}

\acknowledgments

Studies of the evolution of hot gas in elliptical galaxies
at UC Santa Cruz are supported by
NASA grant NAG 5-3060 and NSF grant
AST-9802994 for which we are very grateful.
FB is supported in part by Grant MURST-Cofin 98.

\clearpage

\vskip.1in
\figcaption[aasroundfig1.ps]{
X-ray surface brightness distributions $\Sigma_x(R,z)$ 
(erg cm$^{-2}$ s$^{-1}$) for emission in the 
ROSAT band (0.1 -- 2.4 keV) shown in one 
quadrant of the rotating E2 galaxy at $t = 13$ Gyrs.
Several contours are labeled 
with values of $\log \Sigma_x(R,z)$.
{\it (a):} Rotating cooling flow with no mass dropout 
($q = 0$) viewed perpendicular to the axis of rotation;
contours are evenly spaced
by $\Delta (\log \Sigma_x) = 0.15$.
{\it (b):} Same cooling flow shown in (a) but viewed at 
an inclination of 60$^o$.
{\it (c):} Rotating cooling flow with mass dropout 
($q = 1$) viewed perpendicular to the axis of rotation; 
contours are evenly spaced
by $\Delta (\log \Sigma_x) = 0.3$.
{\it (d):} Same cooling flow shown in (c) but viewed at 
an inclination of 60$^o$.
\label{fig1}}

\vskip.1in
\figcaption[aasroundfig2.ps]{
Azimuthal velocity distribution $v_{\phi}(R,z)$ 
(km s$^{-1}$) in the cooling flow at $t = 13$ Gyrs 
when viewed perpendicular to the axis of rotation. 
{\it (a):} $v_{\phi}(R,z)$ for a cooling flow 
with no mass dropout
($q = 0$); the spacing between adjacent contours is 
$29.4$ km s$^{-1}$, which is also the velocity on the 
contour closest to the $z$-axis.
{\it (b):} $v_{\phi}(R,z)$ for a cooling flow 
with mass dropout
($q = 1$); the spacing between adjacent contours is
$23.5$ km s$^{-1}$, which is also the velocity on the
contour closest to the $z$-axis.
{\it (c):} the emission-weighted line of 
sight velocity $v_{los}(R,z)$ for a cooling flow 
with mass dropout
($q = 1$); the spacing between adjacent contours is
$14.7$ km s$^{-1}$, which is also the velocity on the
contour closest to the $z$-axis.
\label{fig2}}

\vskip.1in
\figcaption[aasroundfig3.ps]{
Comparison of the spatial distribution of cooled 
gas (and young star) 
density $\rho_{do}(R,z)$ ({\it solid contours})
with that of the old stellar population
$\rho_*(R,z)$ ({\it dashed contours}) 
for the $q = 1$ solution at 13 Gyrs.
Contours are labeled with $\log \rho_{do}(R,z)$ 
or $\log \rho_*(R,z)$
with units gm cm$^{-3}$.
The spacing between adjacent 
contours is $\Delta (\log \rho) = 0.35$. 
\label{fig3}}

\vskip.1in
\figcaption[aasroundfig4.ps]{
X-ray surface brightness images $\Sigma_x(R,z)$ 
of the E2 galaxy 
at $t_n = 13$ Gyrs in the ROSAT band 
including emission from mass dropout with $q = 1$. 
Several isophotal contours are labeled with 
$\log \Sigma_x$ in cgs units and contours 
are evenly spaced in $\Delta (\log \Sigma_x) = 0.3$.
{\it (a):} $\Sigma_x(R,z)$ computed with viscosity 
$\mu_{sn} = 16.2 r_{kpc}^{-1.4}$ gm cm$^{-1}$ s$^{-1}$ 
and viewed perpendicular to the axis of rotation.
{\it (b):} $\Sigma_x(R,z)$ computed with viscosity
$\mu_{sn} = 100 r_{kpc}^{-1.4}$ gm cm$^{-1}$ s$^{-1}$ 
and viewed perpendicular to the axis of rotation.
{\it (c):} $\Sigma_x(R,z)$ computed with viscosity
$\mu_{sn} = 100 r_{kpc}^{-1.4}$ gm cm$^{-1}$ s$^{-1}$ 
but viewed at inclination $i = 60^o$. 
\label{fig4}}

\vskip.1in
\figcaption[aasroundfig5.ps]{
ROSAT band 
X-ray surface brightness distributions $\Sigma_x(R,z)$
(erg cm$^{-2}$ s$^{-1}$) in the rotating E2
galaxy viewed perpendicular to the axis of rotation
at $t = 13$ Gyrs.
Several contours are labeled
with values of $\log \Sigma_x(R,z)$ 
and the contours are evenly spaced
by $\Delta (\log \Sigma_x) = 0.3$.
{\it (a):} Computed with uniform viscosity 
$\mu = 5$ gm cm$^{-1}$ s$^{-1}$. 
{\it (b):} Computed with uniform viscosity
$\mu = 20$ gm cm$^{-1}$ s$^{-1}$.
\label{fig5}}

\vskip.1in
\figcaption[aasroundfig6.ps]{
Interstellar iron abundance profiles along the $R$ axis 
({\it solid lines}) and 
the $z$ axis ({\it dashed lines}) at 13 Gyrs 
for the rotating E2 galaxy with $q = 1$ and  
two viscosities:
$\mu = \mu_{sn} = 16.2 r_{kpc}^{-1.4}$ 
({\it heavy lines}) and
uniform $\mu = 100$ ({\it light lines}).
The dotted lines show the stellar iron abundance 
along the $R$ axis ({\it heavy dotted line})
and the $z$ axis ({\it light dotted line}).
\label{fig6}}

\clearpage

\begin{deluxetable}{ll}
\tablewidth{15pc}
\tablenum{1}
\tablecolumns{2}
\tablecaption{GALACTIC PARAMETERS FOR E0 GALAXY}
\tablehead{
\colhead{Parameter} &
\colhead{Value} \cr
}
\startdata
$R_{c*}$        &       311.59 pc\cr
$R_e$\tablenotemark{a}        &       5.088 kpc\cr
$R_t$           &       113.1 kpc\cr
$\beta = R_{ch}/R_{c*}$         &       19.953\cr
$\rho_{*o}$     &       $1.438 \times 10^{-20}$~ gm cm$^{-3}$\cr
$\gamma = \rho_{ho}/\rho_{*o}$  &       $3.7954 \times 10^{-4}$\cr
$M_{*t}$        &       $4.52 \times 10^{11}$ $M_{\odot}$\cr
$M_{ht}$        &       9 $M_{*t}$\cr
$L_B$           &       $4.95 \times 10^{10} L_{B\odot}$\cr
$M_{*t}/L_B$    &       9.14\cr
$\sigma_*$\tablenotemark{b}   &       351 km s$^{-1}$\cr
\enddata
\tablenotetext{a}{Effective radius.}

\tablenotetext{b}{Characteristic velocity dispersion in stellar core,
$\sigma_* = ( 4 \pi G \rho_{*o} R_{c*}^2 /9 )^{1/2}$.}

\end{deluxetable}

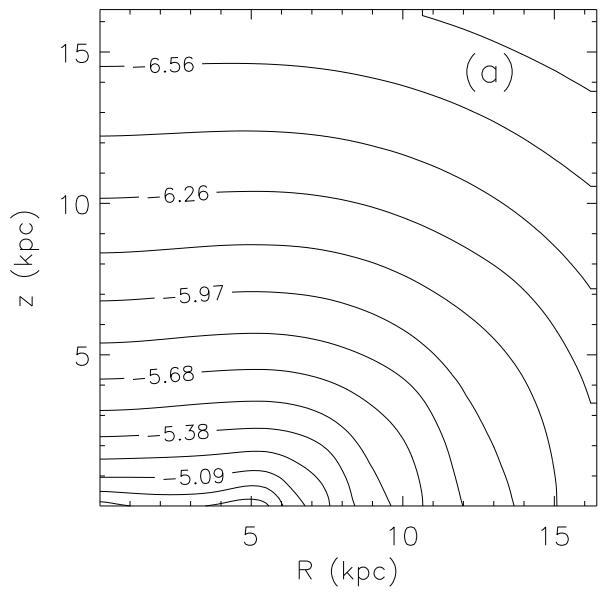
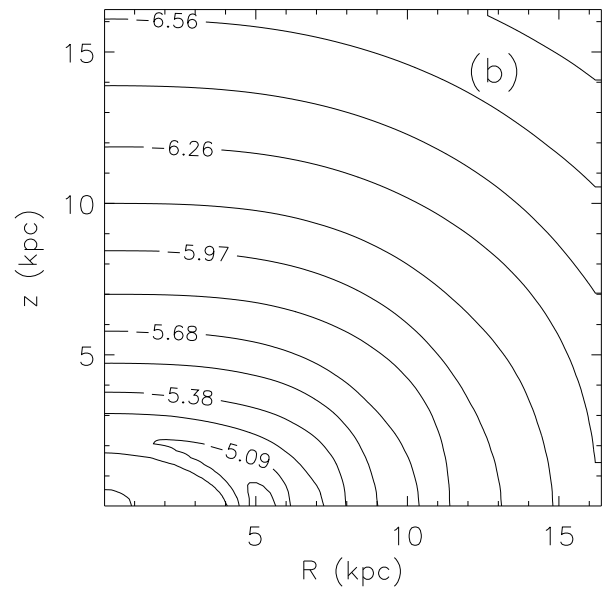
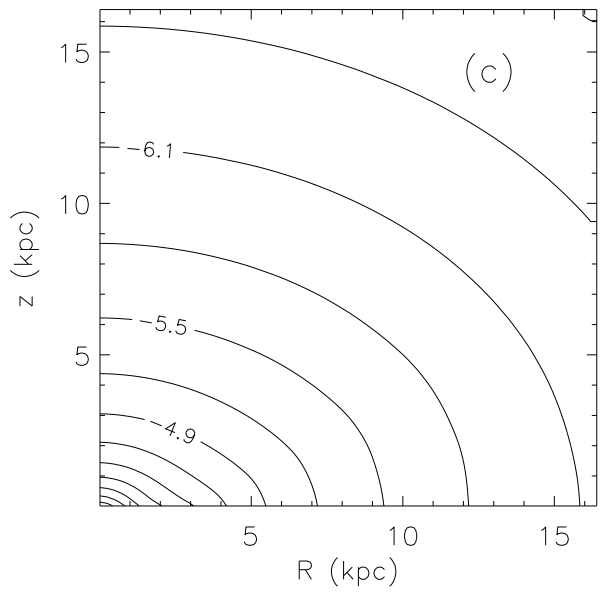
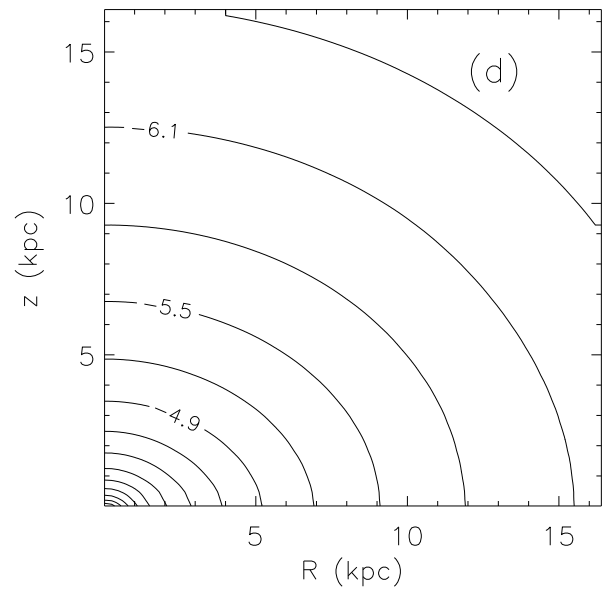
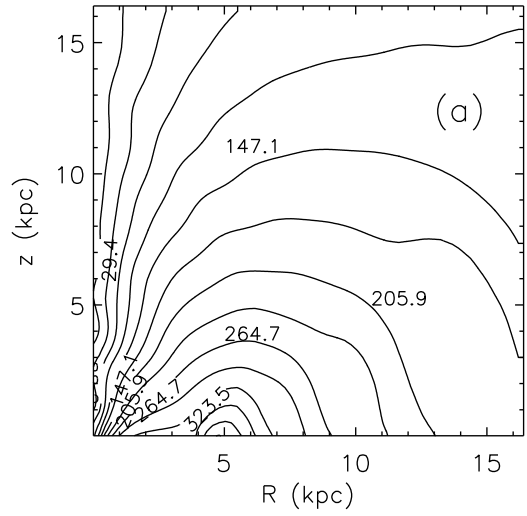
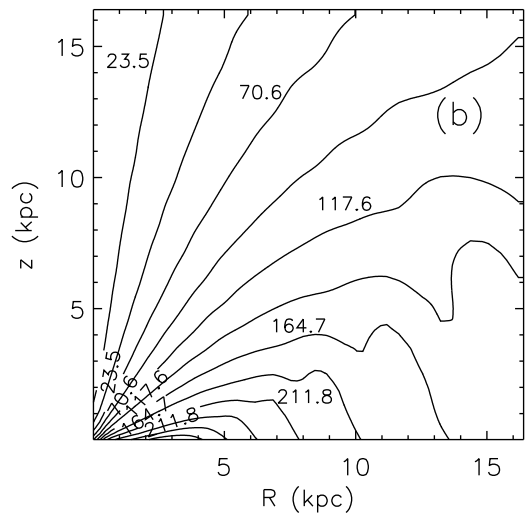
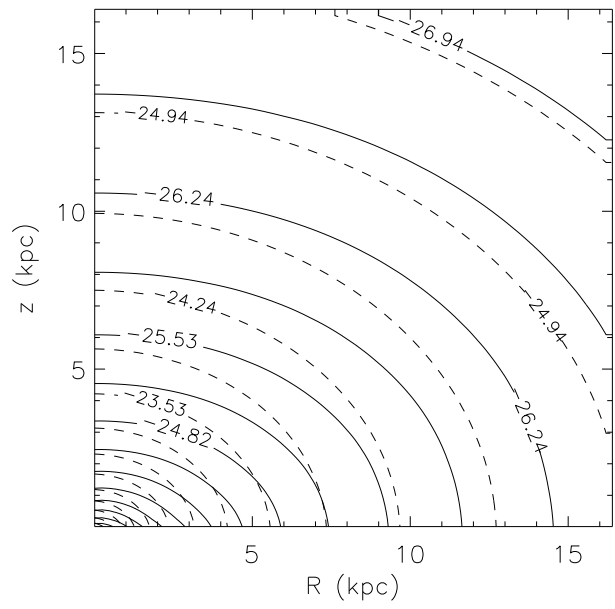
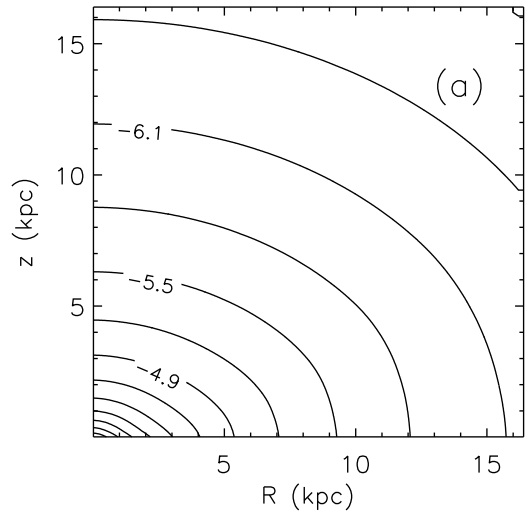
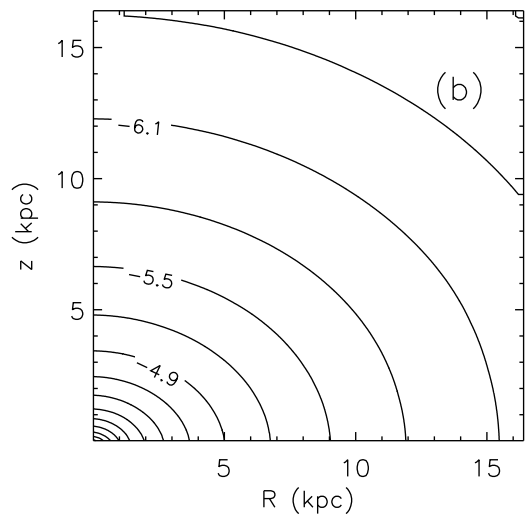
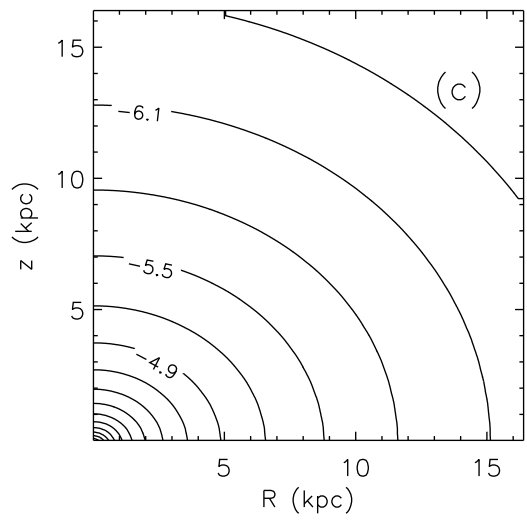
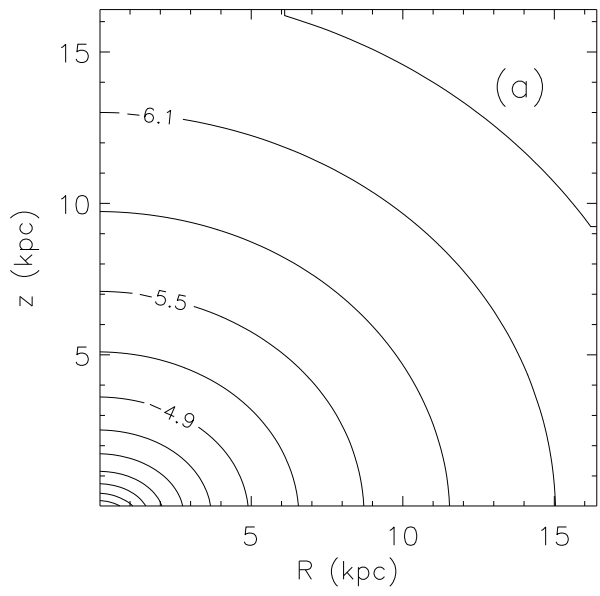
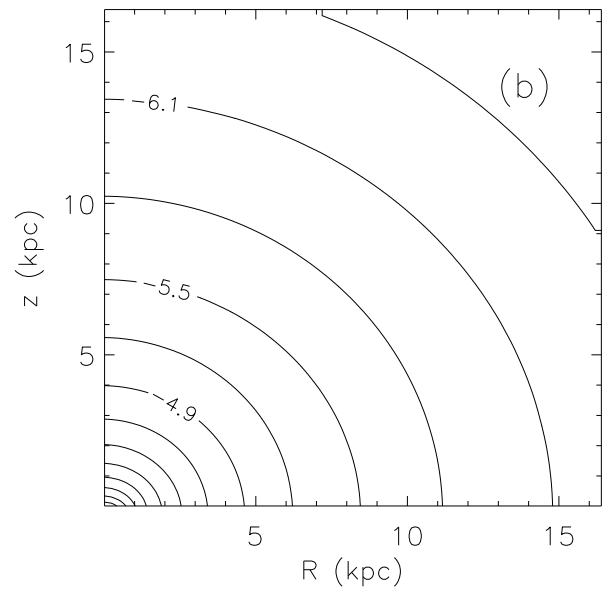
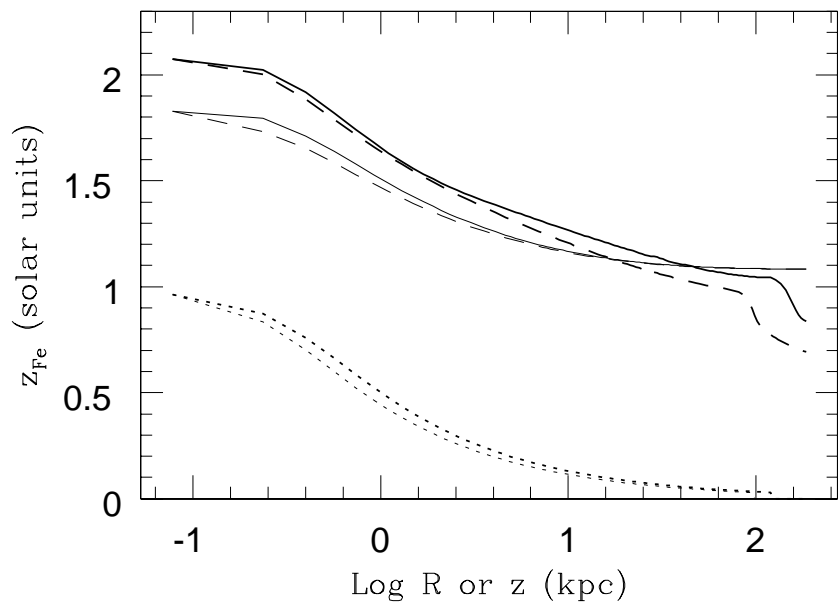
\end{document}